\documentclass{article}

\usepackage{subfigure}
\usepackage{latexsym}
\usepackage{amssymb}
\usepackage{amsthm}
\usepackage{amsmath}
\usepackage{graphics}
\usepackage{graphicx}

\usepackage{url}
\usepackage{listings}
\usepackage{seqsplit}
\usepackage[]{algorithm2e}
\usepackage{hyperref}

\begin{document}

\title{Post-apocalyptic computing from cellular automata}
\author{Genaro J. Mart{\'i}nez$^{1,2,3}$ \& Andrew Adamatzky$^{2,1}$ \& Guanrong Chen$^{3,1}$}

\maketitle

\noindent {\bf Book: {\it Post-Apocalyptic Computing}. WSPC Book Series in Unconventional Computing: Volume 5, Chapter 11, pages 283-296, 2025. URL: \url{https://www.worldscientific.com/worldscibooks/10.1142/13960}} \\

\begin{centering}
$^1$ Artificial Life Robotics Laboratory, Escuela Superior de C\'omputo, Instituto Polit\'ecnico Nacional, M\'exico. \\ \url{gjuarezm@ipn.mx} \\
$^2$ Unconventional Computing Lab, University of the West of England, Bristol, United Kingdom. \\
\url{andrew.adamatzky@uwe.ac.uk} \\
$^3$ Centre for Complexity and Complex Networks, City University of Hong Kong, Hong Kong, China. \\
\url{eegchen@cityu.edu.hk} \\
\end{centering}

\begin{abstract}
Cellular automata are arrays of finite state machines that can exist in a finite number of states. These machines update their states simultaneously based on specific local rules that govern their interactions. This framework provides a simple yet powerful model for studying complex systems and emergent behaviors. We revisit and reconsider the traditional notion of an algorithm, proposing a novel perspective in which algorithms are represented through the dynamic state-space configurations of cellular automata. By doing so, we establish a conceptual framework that connects computation to physical processes in a unique and innovative way. This approach not only enhances our understanding of computation but also paves the way for the future development of unconventional computing devices. Such devices could be engineered to leverage the inherent computational capabilities of physical, chemical, and biological substrates. This opens up new possibilities for designing systems that are more efficient, adaptive, and capable of solving problems in ways that traditional silicon-based computers cannot. The integration of cellular automata into these domains highlights their potential as a transformative tool in the ongoing evolution of computational theory and practice.
\end{abstract}

\section{Introduction}
\begin{figure}[th]
\centerline{\includegraphics[width=11.0cm]{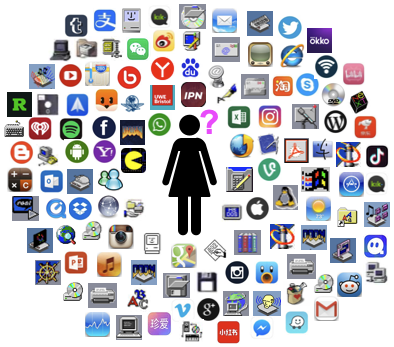}}
\caption{Are you free to make a decision regarding the algorithm to be employed?}
\label{logos}
\end{figure}

How do you decide to choose an algorithm or not? (Fig.~\ref{logos}) This selection could be random, influenced, or inadvertent. When buy a mobile, a number of algorithms pre-installed are running. Consequently, you do not know what is happening with your personal information if you have no experience in computer science. Surely, several of these pre-installed applications upload information to the internet.

Every day, we encounter or discover problems, and we promptly find solutions for some of them. In the realm of computer science, this process involves the development of algorithms. Consequently, the community has generated a number of algorithmic solutions that can be implemented across various devices. Many of these algorithms can be stored in databases such as GitHub, The Algorithms, Reddit, OpenML, and others.

Indeed, some artificial intelligence-based applications scour the internet, extracting code snippets from these repositories to answer queries for users. Examples of such services include ChatGPT, BING chat, Copilot, Bard, or similar platforms. A collapse in our information technology landscape would not signify the commencement of a new era but rather, a challenge to enhance the work of our previous developers.

Several years ago, one of the authors  experienced an anecdote while collaborating with Prof. Harold McIntosh in Puebla. In the midst of a routine maintenance session for his NextStep computer, which involved creating a backup of the hard disk, an unfortunate accident transpired, resulting in the burning of the hard disk. Over ten years of work were stored on that hard disk, and Harold's response was remarkably. Subsequently, he acknowledged the loss of the information and began contemplating how to initiate fresh work—recreating programs, rewriting reports, and so on. This incident presented a potential scenario for starting anew, particularly in a hypothetical post-apocalyptic computing stage.

How can one systematically generate algorithms, especially when considering the essence of computation—the development of programs and algorithms? The systematic and automatic definition of these processes can be both costly and intricate (Fig.~\ref{logos}). This paper explores examples illustrating the intricacy associated with certain codes. Within this framework, we employ cellular automata theory to generate programs based on personal decisions, examining their complexity in a randomised fashion.

Cellular automata field was initiated by John von Neumann in the 1950s. It was inspired by the exploration and resolution of problems such as inherent parallel computing, nervous systems, self-reproduction of machines, and biological cellular processes \cite{von1966theory}. Over more than 60 years of research, this theory has found numerous applications in various academic fields, including biology, physics, economics, mathematics, chemistry, sociology, computer science, patterns, tilings, astronomy, nature, artificial life, complex systems, and so on \cite{toffoli1987cellular, gutowitz1991cellular, wolfram1994cellular, langton1995artificial, sipper1997evolution, adamatzky2001computing, ilachinski2001cellular, wolfram2002new, chopard2005cellular, deutsch2005mathematical, kier2005modeling, margenstern2007cellular, mcintosh2009one, boccara2010modeling, mainzer2011universe, alonso2011discrete, goles2013cellular, morita2017theory, das2023proceedings, zenil2013computable}.

The logic to implement a cellular automaton is simple; however, certain rules have the capability to produce highly complex behaviour, giving rise to intricate patterns during the system's evolution. The most renowned instance is Conway's Game of Life, a binary two-dimensional cellular automaton that evolves with orthogonal and diagonal 1-range neighbours (Moore neighbourhood). Despite the simplicity of the conditions, the resulting behaviour is complex. Over 50 years later, researchers of Life continue to report, construct, and discover new patterns and configurations \cite{games1970fantastic, poundstone1985recursive, adamatzky2010game, johnston2022conway, das2022mathematical}.\footnote{LifeWiki. \url{https://conwaylife.com/wiki}}

Cellular automata typically are studied in one, two, three dimensions. However some studies were done in six, seven, eight dimensions discovering very interesting non-trivial collective behaviour \cite{chate1992collective}. At the same time most complex rules can be designed increasing the number of symbols in its alphabet, they can be explored with two powerful and free versions of software: {\it Golly}\footnote{Golly. \url{https://golly.sourceforge.io/}} and {\it DDLab} \cite{wuensche2011exploring}.\footnote{Discrete Dynamics Lab. \url{http://www.ddlab.com/}}

\begin{figure}[!tbp]
\begin{center}
\subfigure[]{\scalebox{0.227}{\includegraphics{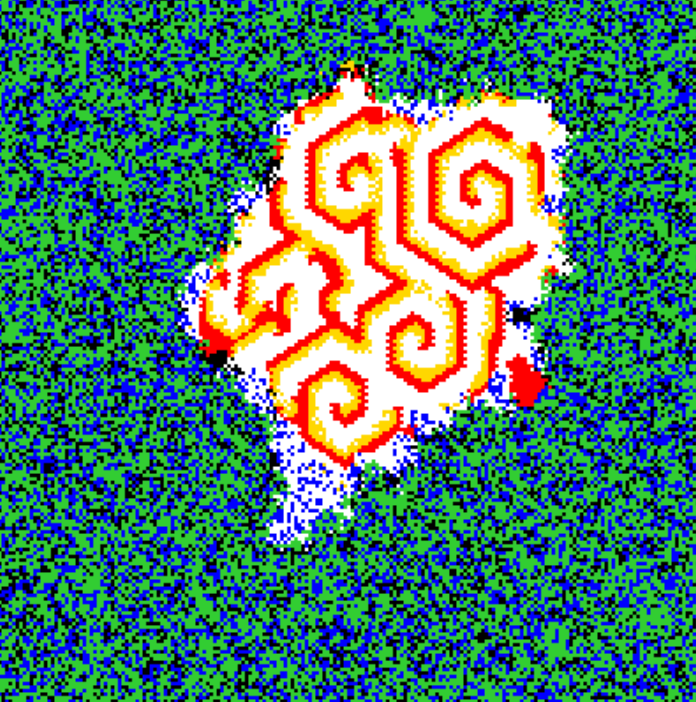}}} %\hspace{0.7cm}
\subfigure[]{\scalebox{0.255}{\includegraphics{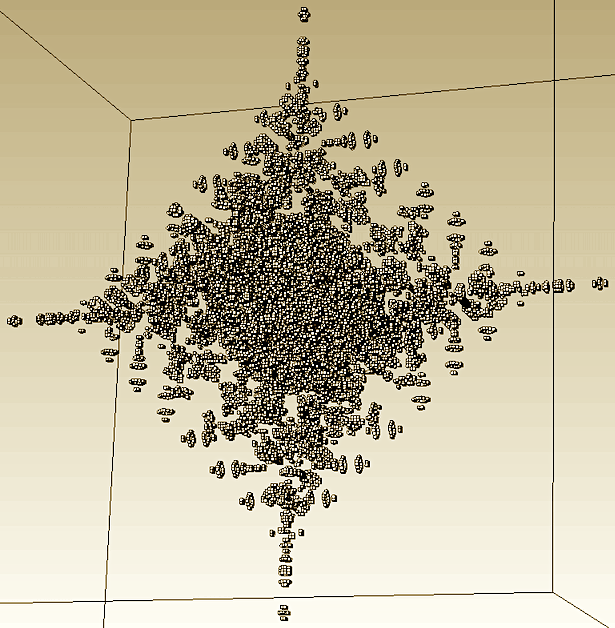}}} %\hspace{0.7cm}
\subfigure[]{\scalebox{0.5}{\includegraphics{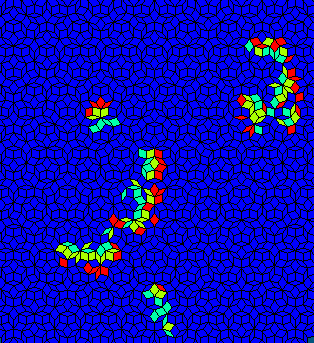}}} %\hspace{0.2cm}
\subfigure[]{\scalebox{0.22}{\includegraphics{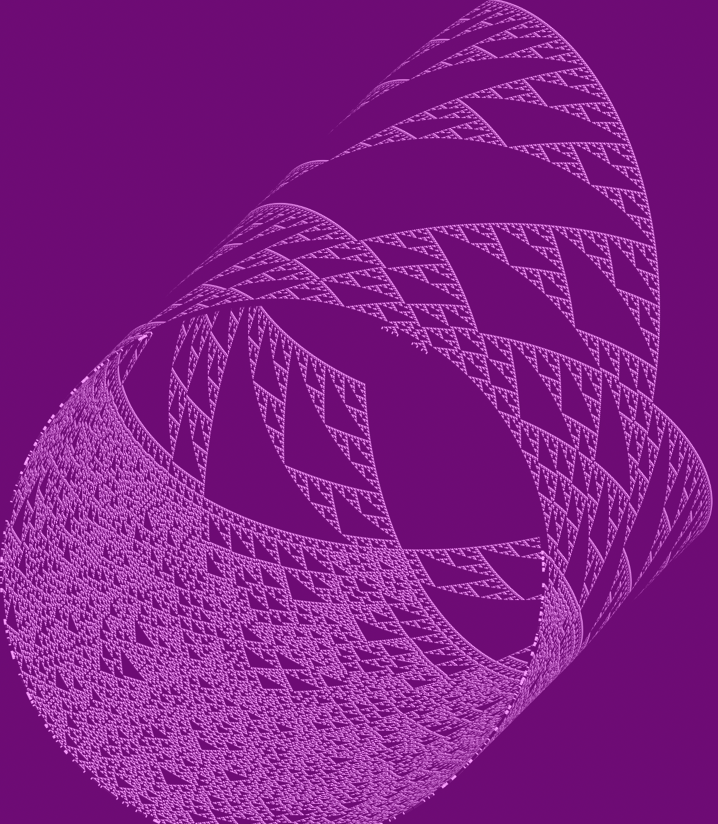}}} %\hspace{0.7cm}
\end{center}
\caption{Examples of cellular automata configurations. (a) Hexagonal lattice with multiple states (DDLab), (b) Three-dimensional space with two states (Ready), (c) Penrose lattice with multiple states (Ready), (d) One-dimensional space with two states (CAViewer).}
\label{kindca}
\end{figure}

Figure~\ref{kindca} displays some kinds of cellular automata. Figure~\ref{kindca}a shows a totalistic two-dimensional hexagonal cellular automaton evolving with six states is calculated the free software DDLab. Figure~\ref{kindca}b shows a semi-totalistic three-dimensional cellular automaton evolving with two states calculated the free software {\it Ready}.\footnote{Ready. \url{https://github.com/GollyGang/ready}} Figure~\ref{kindca}c shows a two-dimensional Penrose cellular automaton evolving with four states calculated the free software {\it Ready}. Figure~\ref{kindca}d shows a binary one-dimensional cellular automaton the elementary evolution rule 22 is calculated the free software {\it CAViewer}.\footnote{CAViewer. \url{https://www.comunidad.escom.ipn.mx/genaro/Cellular_Automata_Repository/Software.html}}

An algorithm is a string. In the 1960s, Marvin Minsky discussed which algorithms have histories of evolution and histories can be drawn in a state machine diagram, as an attractor \cite{minsky1967computation}.

In this manner, we can represent the famous $C$ program \cite{kernighan1988thec}, commonly known as ``the hello world," as a binary string by translating each ASCII symbol of the program into strings of eight symbols. If we were without computers in a post-apocalyptic scenario, the probability of retrieving this program from a binary language would be the order of $2.19 \times 10^{-191}$%.

% 80 characters, 84 B
\begin{center}
\small
\begin{lstlisting}
#include <stdio.h>
int main(void)
{
   printf(''Hello world\n'');
   return 0;
}
\end{lstlisting}
\end{center}

This way, the program {\it hello world} written in $C$ can be expressed in 640 binary symbols given for the next word: \\ 

% 640 symbols, 2^640 = 4562440617622195218641171605700291324893228507248559930579192517899275167208677386505912811317371399778642309573594407310688704721375437998252661319722214188251994674360264950082874192246603776, 4.5624406176221952186411716057002913248932285072485599305791... × 10^192
% (1/(2^640))*100 = 2.191809349008403039752693107141120832630808722124036623272... × 10^-191

{\tiny
\noindent $
\seqsplit{0010001101101001011011100110001101101100011101010110010001100101001000000011110001110011011101000110010001101001011011110010111001101000001111100000101001101001011011100111010000100000011011010110000101101001011011100010100001110110011011110110100101100100001010010000101001111011000010100010000000100000001000000111000001110010011010010110111001110100011001100010100000100111001001110100100001100101011011000110110001101111001000000111011101101111011100100110110001100100010111000110111000100111001001110010100100111011000010100010000000100000001000000111001001100101011101000111010101110010011011100010000000110000001110110000101001111101}
$ \\
}

The next example is a string that encodes a stage of a Fredkin gate in one dimension. Although this construction involves tiling in two dimensions, it operates effectively only in one dimension. This limitation arises from the fact that the memory function requires a few configurations before activation in order to apply the memory. Specifically, it involves the elementary cellular automaton with memory using rule 22 and a majority memory that spans four time steps backward \cite{martinez2018conservative}. It is a chaotic function \cite{chen2018research} with elements of complexity.

A cellular automaton with memory represents an intriguing variant wherein the system has the capacity to recall certain past global states to influence the subsequent state. Conventional cellular automata are ahistorical systems, meaning that the new state of a cell is solely dependent on the neighbourhood configuration at the preceding time step, as dictated by the local function \cite{alonso2011discrete, martinez2013designing}. In this context, cellular automata with memory can be viewed as an extension of the standard framework of classic systems, wherein each cell is granted the ability to remember a certain period of its prior evolution

To implement memory, we need to specify a memory function $\phi$ as follows: $\phi (x^{t-\tau}_{i}, \ldots, x^{t-1}_{i}, x^{t}_{i}) \rightarrow s_{i}$, where $\tau < t$ determines the degree of memory backward, and each cell $s_{i} \in \Sigma$ represents a state function of the series of states of the cell $x_i$ with memory up to the current time step. Finally, to execute the evolution, we apply the original rule: $\varphi(\ldots, s^{t}_{i-1}, s^{t}_{i}, s^{t}_{i+1}, \ldots) \rightarrow x^{t+1}_i$. The key characteristic of cellular automata with memory is that the mapping local function remains unaltered, while the historical memory of all past iterations is retained by representing each cell as a summary of its past states in the memory function.

\begin{figure}[!tbp]
\centerline{\includegraphics[width=12cm]{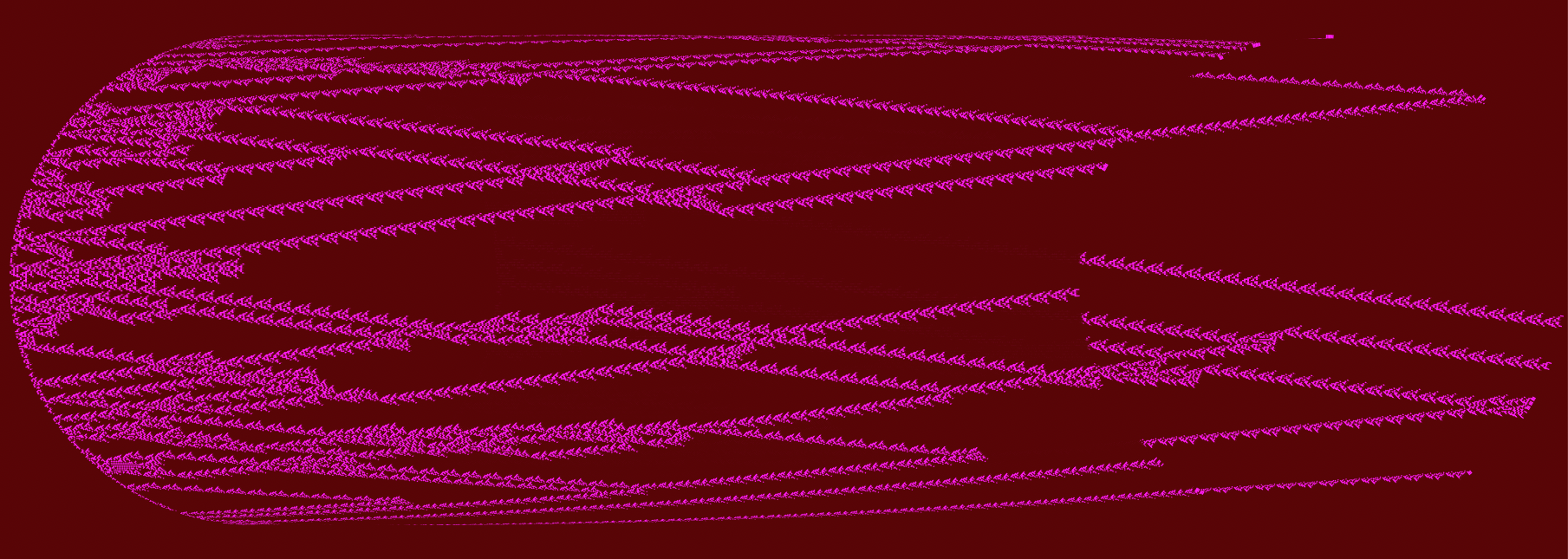}}
\caption{The typical evolution of an elementary cellular automaton with majority memory rule 22 is depicted from a random initial condition. The visualization reveals the emergence of two gliders, traversing and colliding in various scenarios. This snapshot is computed on a tube of 1,200 cells over the course of 615 generations.}
\label{evolR2majm}
\end{figure}

\begin{figure}[!tbp]
\centerline{\includegraphics[width=7.0cm]{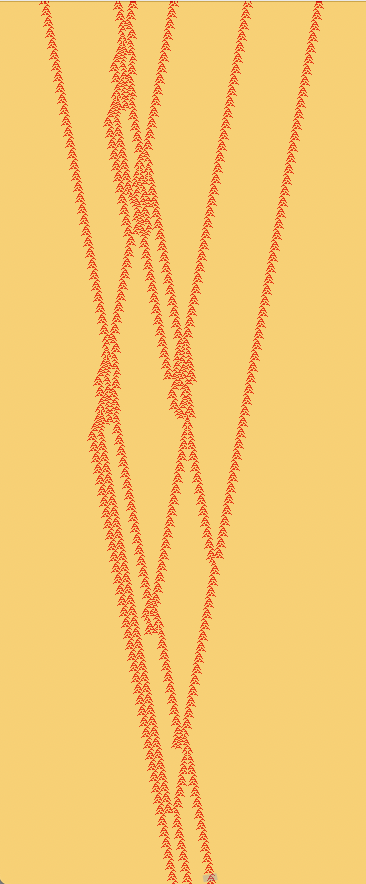}}
\caption{Fredkin gate in elementary cellular automaton with memory rule 22. It is the relation $(c=1, p=1, q=1) \rightarrow (x=1, y=1, z=1)$ running on an initial condition of 413 cells for 882 generations.}
\label{FredkinGateECAM22a}
\end{figure}

The following string represents the initial configuration used to simulate the function of a Fredkin gate in the elementary cellular automaton with memory rule 22 \cite{martinez2011reproducing, barrientos2022synchronizing} (Fig.~\ref{evolR2majm}).\footnote{Soliton collisions in a complex one-dimensional cellular automata with memory. \url{https://youtu.be/qEejQ0KGeTc?si=MD1dBH-fx88o-Nfb}} The string contains 2,065 symbols, resulting in a probability of obtaining this configuration of $2.36 \times 10^{-620}$. A snapshot of this execution is illustrated in Figure~\ref{FredkinGateECAM22a} and the following:

% 2065 symbols, 2^2065 = 4235854619778876348919300317337347943359329425124947923059388635876511598361692195153143583411131313512607640573955017211133238856973146009775394309596549365187330358286876159737881919539072480941972866778276504309175709681925456679644442701132748615920077485118102574581360347256070577665809108057971803491128355506050608977764021015224591252829788705302291684796374143379836265130683310296302694283893423043502880026586232567575131082234481331982001453561713423971401817746416626529095791637552342429817705933131834711742348953957564237508185314608996330444371019095504270740071183627708287968025752844508803397144543232. Numbers = 622 numbers
%
%1/(2^2065)*100 = 2.360798681169569607147356697060545237649960558994269596827... × 10^-620

{\tiny
\noindent $
\seqsplit{0000000000000000000000000000000000000000001111110000000000000000000000000000000000000000000000000000000000000000000000001000000000011100000000000000000000000000000000000000001110000000000000000000000000000000000000000000000000000000000000000000010001000000000000000000000000000000000000000000000000000000000000000000011100000000000000000000000000000000000000000000000000000000000000000000000000000000000000000000000000000000000000000000000000000000000000010000100000000000000000000000000000000000000000000000000000000000000000000000111000000001000100000000000000000000000000000000000000100100000000000000000000000000000000000000000000000000000000000000000000000111000000000000000000000000000000000000000000000000000000000000000001000100000000000000000000000000000000000000000000000000000000000000000000000000000000000000000000000000000000000000000000000000000000000001001011100000000000000000000000000000000000000000000000000000000000000000000010001000000000011100000000000000000000000000000000000011111100000000000000000000000000000000000000000000000000000000000000000000010010000000000000000000000000000000000000000000000000000000000000000000011100000000000000000000000000000000000000000000000000000000000000000000000000000000000000000000000000000000000000000000000000000000000000110001000000000000000000000000000000000000000000000000000000000000000000001110000000000001001000000000000000000000000000000000000100001000000000000000000000000000000000000000000000000000000000000000000001111110000000000000000000000000000000000000000000000000000000000000000001001000000000000000000000000000000000000000000000000000000000000000000000000000000000000000000000000000000000000000000000000000000000000010011001000000000000000000000000000000000000000000000000000000000000000000010010000000000111111000000000000000000000000000000000011101001000000000000000000000000000000000000000000000000000000000000000000010000100000000000000000000000000000000000000000000000000000000000000000111111000000000000000000000000000000000000000000000000000000000000000000000000000000000000000000}
$ \\
}

\begin{table}[!tbp]
\centering
\begin{tabular}{cc}
$\varphi(1,1,1) \rightarrow 0$ & $\varphi(0,1,1) \rightarrow 1$ \\
$\varphi(1,1,0) \rightarrow 1$ & $\varphi(0,1,0) \rightarrow 1$ \\
$\varphi(1,0,1) \rightarrow 1$ & $\varphi(0,0,1) \rightarrow 1$ \\
$\varphi(1,0,0) \rightarrow 0$ & $\varphi(0,0,0) \rightarrow 0$ 
\end{tabular}
%\caption{Local function for Rule 110 - (01101110)$_2$.}
\label{rule110}
\end{table}

The next scenario involves finding a string representing the execution of an algorithm, specifically the initial condition to replicate a cyclic tag system in the elementary cellular automaton rule 110. Rule 110 is a one-dimensional binary cellular automaton in which the local function assesses three variables, defined by equations in Tab.~\ref{rule110}.

\begin{figure}[!tbp]
\centerline{\includegraphics[width=12cm]{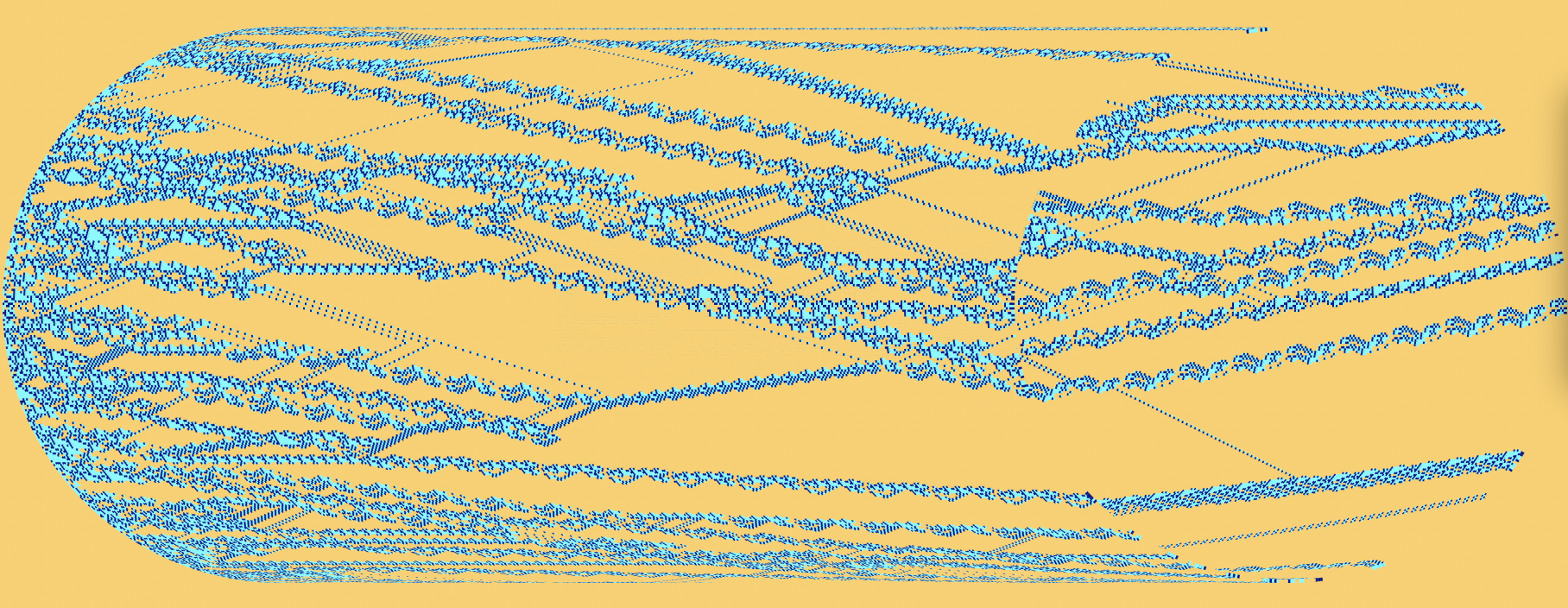}}
\caption{Typical evolution from a random initial condition elementary cellular automaton rule 110 (filtered). It shows the emergence of non-trivial patterns named gliders, travelling and colliding in different scenarios. This snapshot is calculated on a tube of 1,500 cells for 584 generations.}
\label{evolR110}
\end{figure}

\begin{figure}[!tbp]
\centerline{\includegraphics[width=12cm]{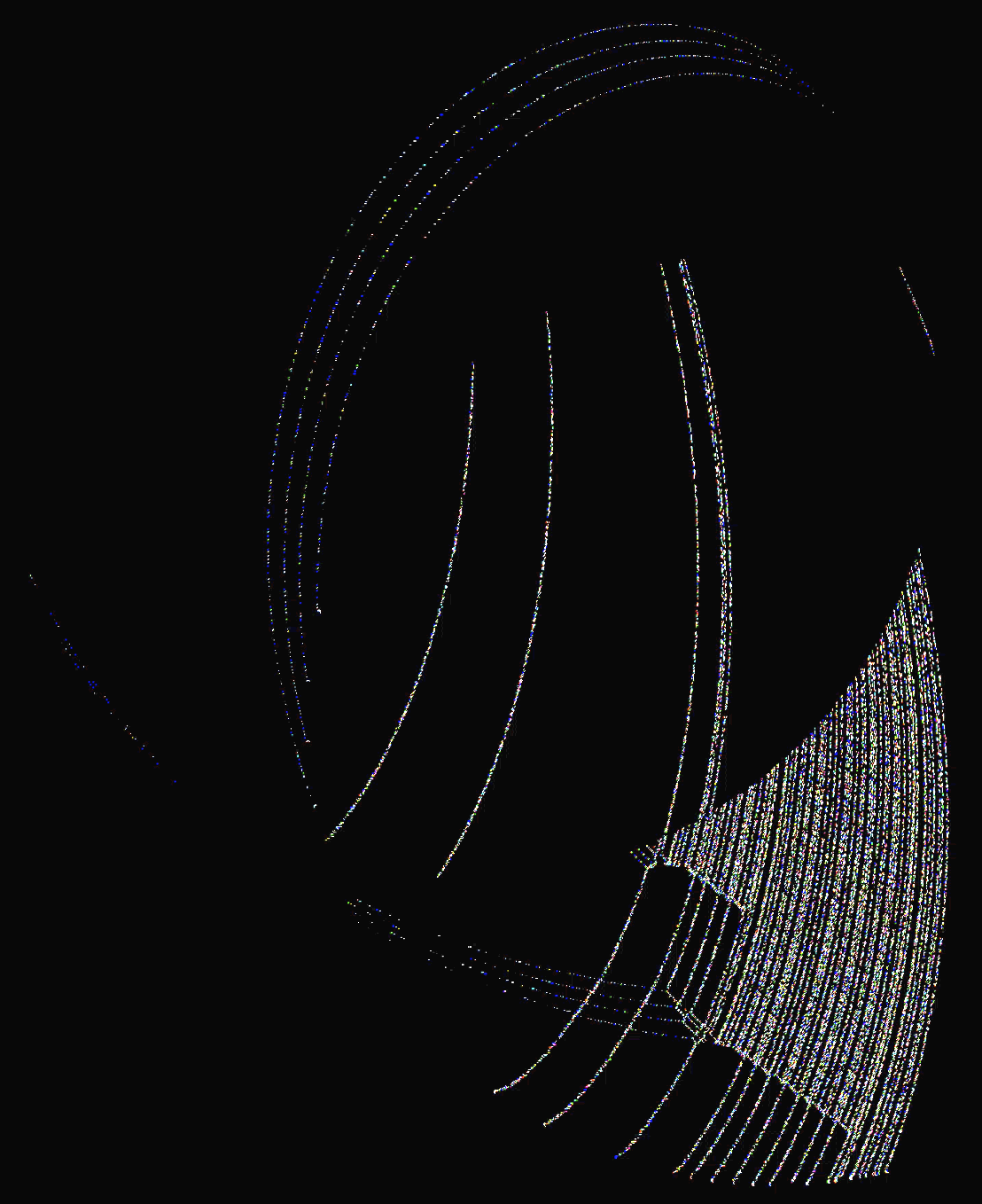}}
\caption{A cyclic tag system evolving in cellular automaton collider elementary rule 110. This snapshot shows the main operations as: read, insert and delete data in the vertical type. To insert new values on the type a lot of particles travel from the left side and they need to cover long distances. This simulation was calculated with DDLab. This simulation implies a space more than $3 \times 10^6$ cells.}
\label{CTSR110-GJMa}
\end{figure}

This function has the ability to produce complex patterns: gliders, particles, waves, mobile self-localisations (Fig.~\ref{evolR110}). These complex patterns emerge during the evolution and the interactions (collisions), determine this artificial universe. A particular interest to study such systems is the capacity to simulate computers in their own dynamics \cite{moore2011nature}. Gliders in rule 110 have a wide variety of types, extensions and combinations. We can handle packages of them in one or several phases. These gliders have important characteristics useful to define distances, slopes, speeds, periods, collisions, and phases \cite{martinez2006gliders}.

Several years ago, a computation was developed in rule 110, implementing a cyclic tag system written by Cook \cite{cook2004universality}. Cyclic tag system is a variant of the classic tag system developed by Emil Post. It is a substitution system that reads the first value of a string, deletes a constant number of them, and inserts new symbols at the end of the string. This way, a cyclic tag system is able to handle more substitutions for every symbol cyclically \cite{cook2004universality, wolfram2002new, martinez2011reproducing, barrientos2022synchronizing}.

The next string represents the initial configuration to simulate the function of a cyclic tag system in the elementary cellular automaton rule 110 \cite{martinez2011reproducing, barrientos2022synchronizing}.\footnote{A Computation in Cellular Automaton Collider Rule 110. \url{https://youtu.be/i5af0tQiVd4?si=2rdQTVF7DsFeSGiz} \\ Virtual Cellular Automata Collider: Two Cyclotrons Connected to a Main Collider. \url{https://youtu.be/cUdRtNwEn9A?si=s8rnkMqOQMtRR_F_}} The string contains 56,240 symbols and therefore the probability to get this configuration is $1.18 \times 10^{-16928}$ (Tab.~\ref{ra_tbl1}). A snapshot of this execution is illustrated in Figure~\ref{CTSR110-GJMa}. \\

%2^56240 = 8.45193488072660818195289565337055100225093848312930091688... × 10^16929
%1/(2^56240) = 1.1834319526627218934911242603550295857988165680473372781... × 10^-16930 (period 78)
% (1/(2^56240))*100 = 1.18 × 10^-16928

{\tiny
\noindent $
\seqsplit{11111000100110111110001001101111100010011011111000100110111110001001101111100010011011111000100110111110001001101111100010011011111000100110111110001001101111100010011011111000100110111110001001101111100010011011111000100110111110001001101111100010011011111000100110111110001001101111100010011011111000100110111110001001101111100010011011111000100110111110001001101111100010011011111000100110111110001001101111100010011011111000100110111110001001101111100010011011111000100110111110001001101111100010011011111000100110111110001001101111100010011011111000100110111110001001101111100010011011111000100110111110001001101111100010011011111000100110111110001001101111100010011011111000100110111110001001101111100010011011111000100110111110001001101111100010011011111000100110111110001001101111100010011011111000100110111110001001101111100010011011111000100110111110001001101111100010011011111000100110111110001001101111100010011011111000100110111110001001101111100010011011111000100110111110001001101111100010011011111000100110111110001001101111100010011011111000100110111110001001101111100010011011111000100110111110001001101111100010011011111000100110111110001001101111100010011011111000100110111110001001101111100010011011111000100110111110001001101111100010011011111000100110111110001001101111100010011011111000100110111110001001101111100010011011111000100110111110001001101111100010011011111000100110111110001001101111100010011011111000100110111110001001101111100010011011111000100110111110001001101111100010011011111000100110111110001001101111100010011011111000100110111110001001101111100010011011111000100110111110001001101111100010011011111000100110111110001001101111100010011011111000100110111110001001101111100010011011111000100110111110001001101111100010011011111000100110111110001001101111100010011011111000100110111110001001101111100010011011111000100110111110001001101111100010011011111000100110111110001001101111100010011011111000100110111110001001101111100010011011111000100110111110001001101111100010011011111000100110111110001001101111100010011011111000100110111110001001101111100010011011111000100110111110001001101111100010011011111000100110111110001001101111100010011011111000100110111110001001101111100010011011111000100110111110001001101111100010011011111000100110111110001001101111100010011011111000100110111110001001101111100010011011111000100110111110001001101111100010011011111000100110111110001001101111100010011011111000100110111110001001101111100010011011111000100110111110001001101111100010011011111000100110111110001001101111100010011011111000100110111110001001101111100010011011111000100110111110001001101111100010011011111000100110111110001001101111100010011011111000100110111110001001101111100010011011111000100110111110001001101111100010011011111000100110111110001001101111100010011011111000100110111110001001101111100010011011111000100110111110001001101111100010011011111000100110111110001001101111100010011011111000100110111110001001101111100010011011111000100110111110001001101111100010011011111000100110111110001001101111100010011011111000100110111110001001101111100010011011111000100110111110001001101111100010011011111000100110111110001001101111100010011011111000100110111110001001101111100010011011111000100110111110001001101111100010011011111000100110111110001001101111100010011011111000100110111110001001101111100010011011111000100110111110001001101111100010011011111000100110111110001001101111100010011011111000100110111110001001101111100010011011111000100110111110001001101111100010011011111000100110111110001001101111100010011011111000100110111110001001101111100010011011111000100110111110001001101111100010011011111000100110111110001001101111100010011011111000100110111110001001101111100010011011111000100110111110001001101111100010011011111000100110111110001001101111100010011011111000100110111110001001101111100010011011111000100110111110001001101111100010011011111000100110111110001001101111100010011011111000100110111110001001101111100010011011111000100110111110001001101111100010011011111000100110111110001001101111100010011011111000100110111110001001101111100010011011111000100110111110001001101111100010011011111000100110111110001001101111100010011011111000100110111110001001101111100010011011111000100110111110001001101111100010011011111000100110111110001001101111100010011011111000100110111110001001101111100010011011111000100110111110001001101111100010011011111000100110111110001001101111100010011011111000100110111110001001101111100010011011111000100110111110001001101111100010011011111000100110111110001001101111100010011011111000100110111110001001101111100010011011111000100110111110001001101111100010011011111000100110111110001001101111100010011011111000100110111110001001101111100010011011111000100110111110001001101111100010011011111000100110111110001001101111100010011011111000100110111110001001101111100010011011111000100110111110001001101111100010011011111000100110111110001001101111100010011011111000100110111110001001101111100010011011111000100110111110001001101111100010011011111000100110111110001001101111100010011011111000100110111110001001101111100010011011111000100110111110001001101111100010011011111000100110111110001001101111100010011011111000100110111110001001101111100010011011111000100110111110001001101111100010011011111000100110111110001001101111100010011011111000100110111110001001101111100010011011111000100110111110001001101111100010011011111000100110111110001001101111100010011011111000100110111110001001101111100010011011111000100110111110001001101111100010011011111000100110111110001001101111100010011011111000100110111110001001101111100010011011111000100110111110001001101111100010011011111000100110111110001001101111100010011011111000100110111110001001101111100010011011111000100110111110001001101111100010011011111000100110111110001001101111100010011011111000100110111110001001101111100010011011111000100110111110001001101111100010011011111000100110111110001001101111100010011011111000100110111110001001101111100010011011111000100110111110001001101111100010011011111000100110111110001001101111100010011011111000100110111110001001101111100010011011111000100110111110001001101111100010011011111000100110111110001001101111100010011011111000100110111110001001101111100010011011111000100110111110001001101111100010011011111000100110111110001001101111100010011011111000100110111110001001101111100010011011111000100110111110001001101111100010011011111000100110111110001001101111100010011011111000100110111110001001101111100010011011111000100110111110001001101111100010011011111000100110111110001001101111100010011011111000100110111110001001101111100010011011111000100110111110001001101111100010011011111000100110111110001001101111100010011011111000100110111110001001101111100010011011111000100110111110001001101111100010011011111000100110111110001001101111100010011011111000100110111110001001101111100010011011111000100110111110001001101111100010011011111000100110111110001001101111100010011011111000100110111110001001101111100010011011111000100110111110001001101111100010011011111000100110111110001001101111100010011011111000100110111110001001101111100010011011111000100110111110001001101111100010011011111000100110111110001001101111100010011011111000100110111110001110111000100110111110001001101111100010011011111000100110111110001001101111100010011011111000100110111110001001101111100010011011111000100110111110001001101111100010011011111000100110111110001001101111100010011011111000100110111110001001101111100010011011111000100110111110001001101111100010011011111000100110111110001001101111100010011011111000100110111110001001101111100010011011111000100110111110111011111000100110111110001001101111100010011011111000100110111110001001101111100010011011111000100110111110001001101111100010011011111000100110111110001001101111100010011011111000100110111110001001101111100010011011111000100110111110001001101111100010011011111000100110111110001001101111100010011011111000100110111110001001101111100010011011101001101111100010011011111000100110111110001001101111100010011011111000100110111110001001101111100010011011111000100110111110001001101111100010011011111000100110111110001001101111100010011011111000100110111110001001101111100010011011111000100110111110001001101111100010011011111000100110111110001001101111100010011011111000100110111110001001101111100010011011111000111011100010011011111000100110111110001001101111100010011011111000100110111110001001101111100010011011111000100110111110001001101111100010011011111000100110111110001001101111100010011011111000100110111110001001101111100010011011111000100110111110001001101111100010011011111000100110111110001001101111100010011011111000100110111110001001101111100010011011111000100110111110001001101111100010011011111000100110111110001001101111100010011011111000100110111110001001101111100010011011111000100110111110001001101111100010011011111000100110111110001001101111100010011011111000100110111110001001101111100010011011111000100110111110001001101111100010011011111000100110111110001001101111100010011011111000100110111110001001101111100010011011111000100110111110001001101111100010011011111000100110111110001001101111100010011011111000100110111110001001101111100010011011111000100110111110001001101111100010011011111000100110111110001001101111100010011011111000100110111110001001101111100010011011111000100110111110001001101111100010011011111000100110111110001001101111100010011011111000100110111110001001101111100010011011111000100110111110001001101111100010011011111000100110111110001001101111100010011011111000100110111110001001101111100010011011111000100110111110001001101111100010011011111000100110111110001001101111100010011011111000100110111110001001101111100010011011111000100110111110001001101111100010011011111000100110111110001001101111100010011011111000100110111110001001101111100010011011111000100110111110001001101111100010011011111000100110111110001001101111100010011011111000100110111110001001101111100010011011111000100110111110001001101111100010011011111000100110111110001001101111100010011011111000100110111110001001101111100010011011111000100110111110001001101111100010011011111000100110111110001001101111100010011011111000100110111110001001101111100010011011111000100110111110001001101111100010011011111000100110111110001001101111100010011011111000100110111110001001101111100010011011111000100110111110001001101111100010011011111000100110111110001001101111100010011011111000100110111110001001101111100010011011111000100110111110001001101111100010011011111000100110111110001001101111100010011011111000100110111110001001101111100010011011111000100110111110001001101111100010011011111000100110111110001001101111100010011011111000100110111110001001101111100010011011111000100110111110001001101111100010011011111000100110111110001001101111100010011011111000100110111110001001101111100010011011111000100110111110001001101111100010011011111000100110111110001001101111100010011011111000100110111110001001101111100010011011111000100110111110001001101111100010011011111000100110111110001001101111100010011011111000100110111110001001101111100010011011111000100110111110001001101111100010011011111000100110111110001001101111100010011011111000100110111110001001101111100010011011111000100110111110001001101111100010011011111000100110111110001001101111100010011011111000100110111110001001101111100010011011111000100110111110001001101111100010011011111000100110111110001001101111100010011011111000100110111110001001101111100010011011111000100110111110001001101111100010011011111000100110111110001001101111100010011011111000100110111110001001101111100010011011111000100110111110001001101111100010011011111000100110111110001001101111100010011011111000100110111110001001101111100010011011111000100110111110001001101111100010011011111000100110111110001001101111100010011011111000100110111110001001101111100010011011111000100110111110001001101111100010011011111000100110111110001001101111100010011011111000100110111110001001101111100010011011111000100110111110001001101111100010011011111000100110111110001001101111100010011011111000100110111110001001101111100010011011111000100110111110001001101111100010011011111000100110111110001001101111100010011011111000100110111110001001101111100010011011111000100110111110001001101111100010011011111000100110111110001001101111100010011011111000100110111110001001101111100010011011111000100110111110001001101111100010011011111000100110111110001001101111100010011011111000100110111110001001101111100010011011111000100110111110001001101111100010011011111000100110111110001001101111100010011011111000100110111110001001101111100010011011111000100110111110001001101111100010011011111000100110111110001001101111100010011011111000100110111110001001101111100010011011111000100110111110001001101111100010011011111000100110111110001001101111100010011011111000100110111110001001101111100010011011111000100110111110001001101111100010011011111000100110111110001001101111100010011011111000100110111110001001101111100010011011111000100110111110001001101111100010011011111000100110111110001001101111100010011011111000100110111110001001101111100010011011111000100110111110001001101111100010011011111000100110111110001001101111100010011011111000100110111110001001101111100010011011111000100110111110001001101111100010011011111000100110111110001001101111100010011011111000100110111110001001101111100010011011111000100110111110001001101111100010011011111000100110111110001001101111100010011011111000100110111110001001101111100010011011111000100110111110001001101111100010011011111000100110111110001001101111100010011011111000100110111110001001101111100010011011111000100110111110001001101111100010011011111000100110111110001001101111100010011011111000100110111110001001101111100010011011111000100110111110001001101111100010011011111000100110111110001001101111100010011011111000100110111110001001101111100010011011111000100110111110001001101111100010011011111000100110111110001001101111100010011011111000100110111110001001101111100010011011111000100110111110001001101111100010011011111000100110111110001001101111100010011011111000100110111110001001101111100010011011111000100110111110001001101111100010011011111000100110111110001001101111100010011011111000100110111110001001101111100010011011111000100110111110001001101111100010011011111000100110111110001001101111100010011011111000100110111110001001101111100010011011111000100110111110001001101111100010011011111000100110111110001001101111100010011011111000100110111110001001101111100010011011111000100110111110001001101111100010011011111000100110111110001001101111100010011011111000100110111110001001101111100010011011111000100110111110001001101111100010011011111000100110111110001001101111100010011011111000100110111110001001101111100010011011111000100110111110001001101111100010011011111000100110111110001001101111100010011011111000100110111110001001101111100010011011111000100110111110001001101111100010011011111000100110111110001001101111100010011011111000100110111110001001101111100010011011111000100110111110001001101111100010011011111000100110111110001001101111100010011011111000100110111110001001101111100010011011111000100110111110001001101111100010011011111000100110111110001001101111100010011011111000100110111110001001101111100010011011111000100110111110001001101111100010011011111000100110111110001001101111100010011011111000100110111110001001101111100010011011111000100110111110001001101111100010011011111000100110111110001001101111100010011011111000100110111110001001101111100010011011111000100110111110001001101111100010011011111000100110111110001001101111100010011011111000100110111110111011111000100110111110001001101111100010011011111000100110111110001001101111100010011011111000100110111110001001101111100010011011111000100110111110001001101111100010011011111000100110111110001001101111100010011011111000100110111110001001101111100010011011111000100110111110001001101111100010011011111000100110111110001001101111100010011011111000100110111110001001101111100010011011111000100110111010011011111000100110111110001001101111100010011011111000100110111110001001101111100010011011111000100110111110001001101111100010011011111000100110111110001001101111100010011011111000100110111110001001101111100010011011111000100110111110001001101111100010011011111000100110111110001001101111100010011011111000100110111110001001101111100011101110001001101111100010011011111000100110111110001001101111100010011011111000100110111110001001101111100010011011111000100110111110001001101111100010011011111000100110111110001001101111100010011011111000100110111110001001101111100010011011111000100110111110001001101111100010011011111000100110111110001001101111100010011011111000100110111110001001101111100010011011111011101111100010011011111000100110111110001001101111100010011011111000100110111110001001101111100010011011111000100110111110001001101111100010011011111000100110111110001001101111100010011011111000100110111110001001101111100010011011111000100110111110001001101111100010011011111000100110111110001001101111100010011011111000100110111110001001101111100010011011111000100110111110001001101111100010011011111000100110111110001001101111100010011011111000100110111110001001101111100010011011111000100110111110001001101111100010011011111000100110111110001001101111100010011011111000100110111110001001101111100010011011111000100110111110001001101111100010011011111000100110111110001001101111100010011011111000100110111110001001101111100010011011111000100110111110001001101111100010011011111000100110111110001001101111100010011011111000100110111110001001101111100010011011111000100110111110001001101111100010011011111000100110111110001001101111100010011011111000100110111110001001101111100010011011111000100110111110001001101111100010011011111000100110111110001001101111100010011011111000100110111110001001101111100010011011111000100110111110001001101111100010011011111000100110111110001001101111100010011011111000100110111110001001101111100010011011111000100110111110001001101111100010011011111000100110111110001001101111100010011011111000100110111110001001101111100010011011111000100110111110001001101111100010011011111000100110111110001001101111100010011011111000100110111110001001101111100010011011111000100110111110001001101111100010011011111000100110111110001001101111100010011011111000100110111110001001101111100010011011111000100110111110001001101111100010011011111000100110111110001001101111100010011011111000100110111110001001101111100010011011111000100110111110001001101111100010011011111000100110111110001001101111100010011011111000100110111110001001101111100010011011111000100110111110001001101111100010011011111000100110111110001001101111100010011011111000100110111110001001101111100010011011111000100110111110001001101111100010011011111000100110111110001001101111100010011011111000100110111110001001101111100010011011111000100110111110001001101111100010011011111000100110111110001001101111100010011011111000100110111110001001101111100010011011111000100110111110001001101111100010011011111000100110111110001001101111100010011011111000100110111110001001101111100010011011111000100110111110001001101111100010011011111000100110111110001001101111100010011011111000100110111110001001101111100010011011111000100110111110001001101111100010011011111000100110111110001001101111100010011011111000100110111110001001101111100010011011111000100110111110001001101111100010011011111000100110111110001001101111100010011011111000100110111110001001101111100010011011111000100110111110001001101111100010011011111000100110111110001001101111100010011011111000100110111110001001101111100010011011111000100110111110001001101111100010011011111000100110111110001001101111100010011011111000100110111110001001101111100010011011111000100110111110001001101111100010011011111000100110111110001001101111100010011011111000100110111110001001101111100010011011111000100110111110001001101111100010011011111000100110111110001001101111100010011011111000100110111110001001101111100010011011111000100110111110001001101111100010011011111000100110111110001001101111100010011011111000100110111110001001101111100010011011111000100110111110001001101111100010011011111000100110111110001001101111100010011011111000100110111110001001101111100010011011111000100110111110001001101111100010011011111000100110111110001001101111100010011011111000100110111110001001101111100010011011111000100110111110001001101111100010011011111000100110111110001001101111100010011011111000100110111110001001101111100010011011111000100110111110001001101111100010011011111000100110111110001001101111100010011011111000100110111110001001101111100010011011111000100110111110001001101111100010011011111000100110111110001001101111100010011011111000100110111110001001101111100010011011111000100110111110001001101111100010011011111000100110111110001001101111100010011011111000100110111110001001101111100010011011111000100110111110001001101111100010011011111000100110111110001001101111100010011011111000100110111110001001101111100010011011111000100110111110001001101111100010011011111000100110111110001001101111100010011011111000100110111110001001101111100010011011111000100110111110001001101111100010011011111000100110111110001001101111100010011011111000100110111110001001101111100010011011111000100110111110001001101111100010011011111000100110111110001001101111100010011011111000100110111110001001101111100010011011111000100110111110001001101111100010011011111000100110111110001001101111100010011011111000100110111110001001101111100010011011111000100110111110001001101111100010011011111000100110111110001001101111100010011011111000100110111110001001101111100010011011111000100110111110001001101111100010011011111000100110111110001001101111100010011011111000100110111110001001101111100010011011111000100110111110001001101111100010011011111000100110111110001001101111100010011011111000100110111110001001101111100010011011111000100110111110001001101111100010011011111000100110111110001001101111100010011011111000100110111110001001101111100010011011111000100110111110001001101111100010011011111000100110111110001001101111100010011011111000100110111110001001101111100010011011111000100110111110001001101111100010011011111000100110111110001001101111100010011011111000100110111110001001101111100010011011111000100110111110001001101111100010011011111000100110111110001001101111100010011011111000100110111110001001101111100010011011111000100110111110001001101111100010011011111000100110111110001001101111100010011011111000100110111110001001101111100010011011111000100110111110001001101111100010011011111000100110111110001001101111100010011011111000100110111110001001101111100010011011111000100110111110001001101111100010011011111000100110111110001001101111100010011011111000100110111110001001101111100010011011111000100110111110001001101111100010011011111000100110111110001001101111100010011011111000100110111110001001101111100010011011111000100110111110001001101111100010011011111000100110111110001001101111100010011011111000100110111110001001101111100010011011111000100110111110001001101111100010011011111000100110111110001001101111100010011011111000100110111110001001101111100010011011111000100110111110001001101111100010011011111000100110111110001001101111100010011011111000100110111110001001101111100010011011111000100110111110001001101111100010011011111000100110111110001001101111100010011011111000100110111110001001101111100010011011111000100110111110001001101111100010011011111000100110111110001001101111100010011011111000100110111110001001101111100010011011111000100110111110001001101111100010011011111000100110111110001001101111100010011011111000100110111110001001101111100010011011111000100110111110001001101111100010011011111000100110111110001001101111100010011011111000100110111110001001101111100010011011111000100110111110001001101111100010011011111000100110111110001001101111100010011011111000100110111110001001101111100010011011111000100110111110001001101111100010011011111000100110111110001001101111100010011011111000100110111010011011111000100110111110001001101111100010011011111000100110111110001001101111100010011011111000100110111110001001101111100010011011111000100110111110001001101111100010011011111000100110111110001001101111100010011011111000100110111110001001101111100010011011111000100110111110001001101111100010011011111000100110111110001001101111100010011011111000100110111110001001101111100010011011111000111011100010011011111000100110111110001001101111100010011011111000100110111110001001101111100010011011111000100110111110001001101111100010011011111000100110111110001001101111100010011011111000100110111110001001101111100010011011111000100110111110001001101111100010011011111000100110111110001001101111100010011011111000100110111110001001101111101110111110001001101111100010011011111000100110111110001001101111100010011011111000100110111110001001101111100010011011111000100110111110001001101111100010011011111000100110111110001001101111100010011011111000100110111110001001101111100010011011111000100110111110001001101111100010011011111000100110111110001001101111100010011011111000100110111110001001101111100010011011101001101111100010011011111000100110111110001001101111100010011011111000100110111110001001101111100010011011111000100110111110001001101111100010011011111000100110111110001001101111100010011011111000100110111110001001101111100010011011111000100110111110001001101111100010011011111000100110111110001001101111100010011011111000100110111110001001101111100010011011111000100110111110001001101111100010011011111000100110111110001001101111100010011011111000100110111110001001101111100010011011111000100110111110001001101111100010011011111000100110111110001001101111100010011011111000100110111110001001101111100010011011111000100110111110001001101111100010011011111000100110111110001001101111100010011011111000100110111110001001101111100010011011111000100110111110001001101111100010011011111000100110111110001001101111100010011011111000100110111110001001101111100010011011111000100110111110001001101111100010011011111000100110111110001001101111100010011011111000100110111110001001101111100010011011111000100110111110001001101111100010011011111000100110111110001001101111100010011011111000100110111110001001101111100010011011111000100110111110001001101111100010011011111000100110111110001001101111100010011011111000100110111110001001101111100010011011111000100110111110001001101111100010011011111000100110111110001001101111100010011011111000100110111110001001101111100010011011111000100110111110001001101111100010011011111000100110111110001001101111100010011011111000100110111110001001101111100010011011111000100110111110001001101111100010011011111000100110111110001001101111100010011011111000100110111110001001101111100010011011111000100110111110001001101111100010011011111000100110111110001001101111100010011011111000100110111110001001101111100010011011111000100110111110001001101111100010011011111000100110111110001001101111100010011011111000100110111110001001101111100010011011111000100110111110001001101111100010011011111000100110111110001001101111100010011011111000100110111110001001101111100010011011111000100110111110001001101111100010011011111000100110111110001001101111100010011011111000100110111110001001101111100010011011111000100110111110001001101111100010011011111000100110111110001001101111100010011011111000100110111110001001101111100010011011111000100110111110001001101111100010011011111000100110111110001001101111100010011011111000100110111110001001101111100010011011111000100110111110001001101111100010011011111000100110111110001001101111100010011011111000100110111110001001101111100010011011111000100110111110001001101111100010011011111000100110111110001001101111100010011011111000100110111110001001101111100010011011111000100110111110001001101111100010011011111000100110111110001001101111100010011011111000100110111110001001101111100010011011111000100110111110001001101111100010011011111000100110111110001001101111100010011011111000100110111110001001101111100010011011111000100110111110001001101111100010011011111000100110111110001001101111100010011011111000100110111110001001101111100010011011111000100110111110001001101111100010011011111000100110111110001001101111100010011011111000100110111110001001101111100010011011111000100110111110001001101111100010011011111000100110111110001001101111100010011011111000100110111110001001101111100010011011111000100110111110001001101111100010011011111000100110111110001001101111100010011011111000100110111110001001101111100010011011111000100110111110001001101111100010011011111000100110111110001001101111100010011011111000100110111110001001101111100010011011111000100110111110001001101111100010011011111000100110111110001001101111100010011011111000100110111110001001101111100010011011111000100110111110001001101111100010011011111000100110111110001001101111100010011011111000100110111110001001101111100010011011111000100110111110001001101111100010011011111000100110111110001001101111100010011011111000100110111110001001101111100010011011111000100110111110001001101111100010011011111000100110111110001001101111100010011011111000100110111110001001101111100010011011111000100110111110001001101111100010011011111000100110111110001001101111100010011011111000100110111110001001101111100010011011111000100110111110001001101111100010011011111000100110111110001001101111100010011011111000100110111110001001101111100010011011111000100110111110001001101111100010011011111000100110111110001001101111100010011011111000100110111110001001101111100010011011111000100110111110001001101111100010011011111000100110111110001001101111100010011011111000100110111110001001101111100010011011111000100110111110001001101111100010011011111000100110111110001001101111100010011011111000100110111110001001101111100010011011111000100110111110001001101111100010011011111000100110111110001001101111100010011011111000100110111110001001101111100010011011111000100110111110001001101111100010011011111000100110111110001001101111100010011011111000100110111110001001101111100010011011111000100110111110001001101111100010011011111000100110111110001001101111100010011011111000100110111110001001101111100010011011111000100110111110001001101111100010011011111000100110111110001001101111100010011011111000100110111110001001101111100010011011111000100110111110001001101111100010011011111000100110111110001001101111100010011011111000100110111110001001101111100010011011111000100110111110001001101111100010011011111000100110111110001001101111100010011011111000100110111110001001101111100010011011111000100110111110001001101111100010011011111000100110111110001001101111100010011011111000100110111110001001101111100010011011111000100110111110001001101111100010011011111000100110111110001001101111100010011011111000100110111110001001101111100010011011111000100110111110001001101111100010011011111000100110111110001001101111100010011011111000100110111110001001101111100010011011111000100110111110001001101111100010011011111000100110111110001001101111100010011011111000100110111110001001101111100010011011111000100110111110001001101111100010011011111000100110111110001001101111100010011011111000100110111110001001101111100010011011111000100110111110001001101111100010011011111000100110111110001001101111100010011011111000100110111110001001101111100010011011111000100110111110001001101111100010011011111000100110111110001001101111100010011011111000100110111110001001101111100010011011111000100110111110001001101111100010011011111000100110111110001001101111100010011011111000100110111110001001101111100010011011111000100110111110001001101111100010011011111000100110111110001001101111100010011011111000100110111110001001101111100010011011111000100110111110001001101111100010011011111000100110111110001001101111100010011011111000100110111110001001101111100010011011111000100110111110001001101111100010011011111000100110111110001001101111100010011011111000100110111110001001101111100010011011111000100110111110001001101111100010011011111000100110111110001001101111100010011011111000100110111110001001101111100010011011111000100110111110001001101111100010011011111000100110111110001001101111100010011011111000100110111110001001101111100010011011111000100110111110001001101111100010011011111000100110111110001001101111100010011011111000100110111110001001101111100010011011111000100110111110001001101111100010011011111000100110111110001001101111100010011011111000111011100010011011111000100110111110001001101111100010011011111000100110111110001001101111100010011011111000100110111110001001101111100010011011111000100110111110001001101111100010011011111000100110111110001001101111100010011011111000100110111110001001101111100010011011111000100110111110001001101111100010011011111000100110111110001001101111100010011011111000100110111110001001101111100010011011111011101111100010011011111000100110111110001001101111100010011011111000100110111110001001101111100010011011111000100110111110001001101111100010011011111000100110111110001001101111100010011011111000100110111110001001101111100010011011111000100110111110001001101111100010011011111000100110111110001001101111100010011011111000100110111110001001101110100110111110001001101111100010011011111000100110111110001001101111100010011011111000100110111110001001101111100010011011111000100110111110001001101111100010011011111000100110111110001001101111100010011011111000100110111110001001101111100010011011111000100110111110001001101111100010011011111000100110111110001001101111100010011011111000100110111110001001101111100011101110001001101111100010011011111000100110111110001001101111100010011011111000100110111110001001101111100010011011111000100110111110001001101111100010011011111000100110111110001001101111100010011011111000100110111110001001101111100010011011111000100110111110001001101111100010011011111000100110111110001001101111100010011011111000100110111110001001101111100010011011111000100110111110001001101111100010011011111000100110111110001001101111100010011011111000100110111110001001101111100010011011111000100110111110001001101111100010011011111000100110111110001001101111100010011011111000100110111110001001101111100010011011111000100110111110001001101111100010011011111000100110111110001001101111100010011011111000100110111110001001101111100010011011111000100110111110001001101111100010011011111000100110111110001001101111100010011011111000100110111110001001101111100010011011111000100110111110001001101111100010011011111000100110111110001001101111100010011011111000100110111110001001101111100010011011111000100110111110001001101111100010011011111000100110111110001001101111100010011011111000100110111110001001101111100010011011111000100110111110001001101111100010011011111000100110111110001001101111100010011011111000100110111110001001101111100010011011111000100110111110001001101111100010011011111000100110111110001001101111100010011011111000100110111110001001101111100010011011111000100110111110001001101111100010011011111000100110111110001001101111100010011011111000100110111110001001101111100010011011111000100110111110001001101111100010011011111000100110111110001001101111100010011011111000100110111110001001101111100010011011111000100110111110001001101111100010011011111000100110111110001001101111100010011011111000100110111110001001101111100010011011111000100110111110001001101111100010011011111000100110111110001001101111100010011011111000100110111110001001101111100010011011111000100110111110001001101111100010011011111000100110111110001001101111100010011011111000100110111110001001101111100010011011111000100110111110001001101111100010011011111000100110111110001001101111100010011011111000100110111110001001101111100010011011111000100110111110001001101111100010011011111000100110111110001001101111100010011011111000100110111110001001101111100010011011111000100110111110001001101111100010011011111000100110111110001001101111100010011011111000100110111110001001101111100010011011111000100110111110001001101111100010011011111000100110111110001001101111100010011011111000100110111110001001101111100010011011111000100110111110001001101111100010011011111000100110111110001001101111100010011011111000100110111110001001101111100010011011111000100110111110001001101111100010011011111000100110111110001001101111100010011011111000100110111110001001101111100010011011111000100110111110001001101111100010011011111000100110111110001001101111100010011011111000100110111110001001101111100010011011111000100110111110001001101111100010011011111000100110111110001001101111100010011011111000100110111110001001101111100010011011111000100110111110001001101111100010011011111000100110111110001001101111100010011011111000100110111110001001101111100010011011111000100110111110001001101111100010011011111000100110111110001001101111100010011011111000100110111110001001101111100010011011111000100110111110001001101111100010011011111000100110111110001001101111100010011011111000100110111110001001101111100010011011111000100110111110001001101111100010011011111000100110111110001001101111100010011011111000100110111110001001101111100010011011111000100110111110001001101111100010011011111000100110111110001001101111100010011011111000100110111110001001101111100010011011111000100110111110001001101111100010011011111000100110111110001001101111100010011011111000100110111110001001101111100010011011111000100110111110001001101111100010011011111000100110111110001001101111100010011011111000100110111110001001101111100010011011111000100110111110001001101111100010011011111000100110111110001001101111100010011011111000100110111110001001101111100010011011111000100110111110001001101111100010011011111000100110111110001001101111100010011011111000100110111110001001101111100010011011111000100110111110001001101111100010011011111000100110111110001001101111100010011011111000100110111110001001101111100010011011111000100110111110001001101111100010011011111000100110111110001001101111100010011011111000100110111110001001101111100010011011111000100110111110001001101111100010011011111000100110111110001001101111100010011011111000100110111110001001101111100010011011111000100110111110001001101111100010011011111000100110111110001001101111100010011011111000100110111110001001101111100010011011111000100110111110001001101111100010011011111000100110111110001001101111100010011011111000100110111110001001101111100010011011111000100110111110001001101111100010011011111000100110111110001001101111100010011011111000100110111110001001101111100010011011111000100110111110001001101111100010011011111000100110111110001001101111100010011011111000100110111110001001101111100010011011111000100110111110001001101111100010011011111000100110111110001001101111100010011011111000100110111110001001101111100010011011111000100110111110001001101111100010011011111000100110111110001001101111100010011011111000100110111110001001101111100010011011111000100110111110001001101111100010011011111000100110111110001001101111100010011011111000100110111110001001101111100010011011111000100110111110001001101111100010011011111000100110111110001001101111100010011011111000100110111110001001101111100010011011111000100110111110001001101111100010011011111000100110111110001001101111100010011011111000100110111110001001101111100010011011111000100110111110001001101111100010011011111000100110111110001001101111100010011011111000100110111110001001101111100010011011111000100110111110001001101111100010011011111000100110111110001001101111100010011011111000100110111110001001101111100010011011111000100110111110001001101111100010011011111000100110111110001001101111100010011011111000100110111110001001101111100010011011111000100110111110001001101111100010011011111000100110111110001001101111100010011011111000100110111110001001101111100010011011111000100110111110001001101111100010011011111000100110111110001001101111100010011011111000100110111110001001101111100010011011111000100110111110001001101111100010011011111000100110111110001001101111100010011011111000100110111110001001101111100010011011111000100110111110001001101111100010011011111000100110111110001001101111100010011011111000100110111110001001101111100010011011111000100110111110001001101111100010011011111000100110111110001001101111100010011011111000100110111110001001101111100010011011111000100110111110001001101111100010011011111000100110111110001001101111100010011011111000100110111110001001101111100010011011111000100110111110001001101111100010011011111000100110111110001001101111100010011011111000100110111110001001101111100010011011111000100110111110001001101111100010011011111000100110111110001001101111100010011011111000100110111110001001101111100010011011111000100110111110001001101111100010011011111000100110111110001001101111100010011011111000100110111110001001101111100010011011111000100110111110001001101111100010011011111000100110111110001001101111100010011011111011101111100010011011111000100110111110001001101111100010011011111000100110111110001001101111100010011011111000100110111110001001101111100010011011111000100110111110001001101111100010011011111000100110111110001001101111100010011011111000100110111110001001101111100010011011111000100110111110001001101111100010011011111000100110111110001001101111100010011011111000100110111110001001101111100010011011101001101111100010011011111000100110111110001001101111100010011011111000100110111110001001101111100010011011111000100110111110001001101111100010011011111000100110111110001001101111100010011011111000100110111110001001101111100010011011111000100110111110001001101111100010011011111000100110111110001001101111100010011011111000100110111110001110111000100110111110001001101111100010011011111000100110111110001001101111100010011011111000100110111110001001101111100010011011111000100110111110001001101111100010011011111000100110111110001001101111100010011011111000100110111110001001101111100010011011111000100110111110001001101111100010011011111000100110111110001001101111100010011011111000100110111110001001101111101110111110001001101111100010011011111000100110111110001001101111100010011011111000100110111110001001101111100010011011111000100110111110001001101111100010011011111000100110111110001001101111100010011011111000100110111110001001101111100010011011111000100110111110001001101111100010011011111000100110111110001001101111100010011011111000100110111110001001101111100010011011111000100110111110001001101111100010011011111000100110111110001001101111100010011011111000100110111110001001101111100010011011111000100110111110001001101111100010011011111000100110111110001001101111100010011011111000100110111110001001101111100010011011111000100110111110001001101111100010011011111000100110111110001001101111100010011011111000100110111110001001101111100010011011111000100110111110001001101111100010011011111000100110111110001001101111100010011011111000100110111110001001101111100010011011111000100110111110001001101111100010011011111000100110111110001001101111100010011011111000100110111110001001101111100010011011111000100110111110001001101111100010011011111000100110111110001001101111100010011011111000100110111110001001101111100010011011111000100110111110001001101111100010011011111000100110111110001001101111100010011011111000100110111110001001101111100010011011111000100110111110001001101111100010011011111000100110111110001001101111100010011011111000100110111110001001101111100010011011111000100110111110001001101111100010011011111000100110111110001001101111100010011011111000100110111110001001101111100010011011111000100110111110001001101111100010011011111000100110111110001001101111100010011011111000100110111110001001101111100010011011111000100110111110001001101111100010011011111000100110111110001001101111100010011011111000100110111110001001101111100010011011111000100110111110001001101111100010011011111000100110111110001001101111100010011011111000100110111110001001101111100010011011111000100110111110001001101111100010011011111000100110111110001001101111100010011011111000100110111110001001101111100010011011111000100110111110001001101111100010011011111000100110111110001001101111100010011011111000100110111110001001101111100010011011111000100110111110001001101111100010011011111000100110111110001001101111100010011011111000100110111110001001101111100010011011111000100110111110001001101111100010011011111000100110111110001001101111100010011011111000100110111110001001101111100010011011111000100110111110001001101111100010011011111000100110111110001001101111100010011011111000100110111110001001101111100010011011111000100110111110001001101111100010011011111000100110111110001001101111100010011011111000100110111110001001101111100010011011111000100110111110001001101111100010011011111000100110111110001001101111100010011011111000100110111110001001101111100010011011111000100110111110001001101111100010011011111000100110111110001001101111100010011011111000100110111110001001101111100010011011111000100110111110001001101111100010011011111000100110111110001001101111100010011011111000100110111110001001101111100010011011111000100110111110001001101111100010011011111000100110111110001001101111100010011011111000100110111110001001101111100010011011111000100110111110001001101111100010011011111000100110111110001001101111100010011011111000100110111110001001101111100010011011111000100110111110001001101111100010011011111000100110111110001001101111100010011011111000100110111110001001101111100010011011111000100110111110001001101111100010011011111000100110111110001001101111100010011011111000100110111110001001101111100010011011111000100110111110001001101111100010011011111000100110111110001001101111100010011011111000100110111110001001101111100010011011111000100110111110001001101111100010011011111000100110111110001001101111100010011011111000100110111110001001101111100010011011111000100110111110001001101111100010011011111000100110111110001001101111100010011011111000100110111110001001101111100010011011111000100110111110001001101111100010011011111000100110111110001001101111100010011011111000100110111110001001101111100010011011111000100110111110001001101111100010011011111000100110111110001001101111100010011011111000100110111110001001101111100010011011111000100110111110001001101111100010011011111000100110111110001001101111100010011011111000100110111110001001101111100010011011111000100110111110001001101111100010011011111000100110111110001001101111100010011011111000100110111110001001101111100010011011111000100110111110001001101111100010011011111000100110111110001001101111100010011011111000100110111110001001101111100010011011111000100110111110001001101111100010011011111000100110111110001001101111100010011011111000100110111110001001101111100010011011111000100110111110001001101111100010011011111000100110111110001001101111100010011011111000100110111110001001101111100010011011111000100110111110001001101111100010011011111000100110111110001001101111100010011011111000100110111110001001101111100010011011111000100110111110001001101111100010011011111000100110111110001001101111100010011011111000100110111110001001101111100010011011111000100110111110001001101111100010011011111000100110111110001001101111100010011011111000100110111110001001101111100010011011111000100110111110001001101111100010011011111000100110111110001001101111100010011011111000100110111110001001101111100010011011111000100110111110001001101111100010011011111000100110111110001001101111100010011011111000100110111110001001101111100010011011111000100110111110001001101111100010011011111000100110111110001001101111100010011011111000100110111110001001101111100010011011111000100110111110001001101111100010011011111000100110111110001001101111100010011011111000100110111110001001101111100010011011111000100110111110001001101111100010011011111000100110111110001001101111100010011011111000100110111110001001101111100010011011111000100110111110001001101111100010011011111000100110111110001001101111100010011011111000100110111110001001101111100010011011111000100110111110001001101111100010011011111000100110111110001001101111100010011011111000100110111110001001101111100010011011111000100110111110001001101111100010011011111000100110111110001001101111100010011011111000100110111110001001101111100010011011111000100110111110001001101111100010011011111000100110111110001001101111100010011011111000100110111110001001101111100010011011111000100110111110001001101111100010011011111000100110111110001001101111100010011011111000100110111110001001101111100010011011111000100110111110001001101111100010011011111000100110111110001001101111100010011011111000100110111110001001101111100010011011111000100110111110001001101111100010011011111000100110111110001001101111100010011011111000100110111110001001101111100010011011111000100110111110001001101111100010011011111000100110111110001001101111100010011011111000100110111110001001101111100010011011111000100110111110001001101111100010011011111000100110111110001001101111100010011011111000100110111110001001101111100010011011111000100110111110001001101111100010011011111000100110111110001001101111100010011011111000100110111110001001101111100010011011111000100110111110001001101111100010011011111000100110111110001001101111100010011011111000100110111110001001101111100010011011111000100110111110001001101111100010011011111000100110111110001001101111100010011011101001101111100010011011111000100110111110001001101111100010011011111000100110111110001001101111100010011011111000100110111110001001101111100010011011111000100110111110001001101111100010011011111000100110111110001001101111100010011011111000100110111110001001101111100010011011111000100110111110001001101111100010011011111000100110111110001001101111100010011011111000100110111110001001101111100011101110001001101111100010011011111000100110111110001001101111100010011011111000100110111110001001101111100010011011111000100110111110001001101111100010011011111000100110111110001001101111100010011011111000100110111110001001101111100010011011111000100110111110001001101111100010011011111000100110111110001001101111100010011011111000100110111110111011111000100110111110001001101111100010011011111000100110111110001001101111100010011011111000100110111110001001101111100010011011111000100110111110001001101111100010011011111000100110111110001001101111100010011011111000100110111110001001101111100010011011111000100110111110001001101111100010011011111000100110111110001001101111100010011011111000100110111110001001101110100110111110001001101111100010011011111000100110111110001001101111100010011011111000100110111110001001101111100010011011111000100110111110001001101111100010011011111000100110111110001001101111100010011011111000100110111110001001101111100010011011111000100110111110001001101111100010011011111000100110111110001001101111100010011011111000100110111110001001101111100010011011111000100110111110001001101111100010011011111000100110111110001001101111100010011011111000100110111110001001101111100010011011111000100110111110001001101111100010011011111000100110111110001001101111100010011011111000100110111110001001101111100010011011111000100110111110001001101111100010011011111000100110111110001001101111100010011011111000100110111110001001101111100010011011111000100110111110001001101111100010011011111000100110111110001001101111100010011011111000100110111110001001101111100010011011111000100110111110001001101111100010011011111000100110111110001001101111100010011011111000100110111110001001101111100010011011111000100110111110001001101111100010011011111000100110111110001001101111100010011011111000100110111110001001101111100010011011111000100110111110001001101111100010011011111000100110111110001001101111100010011011111000100110111110001001101111100010011011111000100110111110001001101111100010011011111000100110111110001001101111100010011011111000100110111110001001101111100010011011111000100110111110001001101111100010011011111000100110111110001001101111100010011011111000100110111110001001101111100010011011111000100110111110001001101111100010011011111000100110111110001001101111100010011011111000100110111110001001101111100010011011111000100110111110001001101111100010011011111000100110111110001001101111100010011011111000100110111110001001101111100010011011111000100110111110001001101111100010011011111000100110111110001001101111100010011011111000100110111110001001101111100010011011111000100110111110001001101111100010011011111000100110111110001001101111100010011011111000100110111110001001101111100010011011111000100110111110001001101111100010011011111000100110111110001001101111100010011011111000100110111110001001101111100010011011111000100110111110001001101111100010011011111000100110111110001001101111100010011011111000100110111110001001101111100010011011111000100110111110001001101111100010011011111000100110111110001001101111100010011011111000100110111110001001101111100010011011111000100110111110001001101111100000010011011111000100110111110001001101111100000010011011111000100110111110001001101111100000010011011111000100110111110001110110101111100010011011111011100010011011111000100110010001110011000111001101111100010011010011111110101111100010011011111000100110111110001001101111100010011001000111001100011111000100110111110001001100111011100110001001101111100010011010001101001100010011011111000100110111110001001101100010111110001001101111100111110111001101111100010011011111000100110111011011101011100110111111001111000100110111110001110111110100111110001001101111100010011011100011101100010011011111000101101000011011111000100110111110001001101111110000110111001101111100010011011111000100110111110011100000100110111011011101011100110111110001001101111100010011011101110011000010011011111000101101000011011111011111101111101011111000100110111110001001101111100111110111001101111100010011011111000100110111011011101011100110111110001001100100011100110001110011011111000100110100111111101011111000100110111110001001101111100010011011111000100110010001110011000111110001001101111100010011001110111001100010011011111000100110100011010011000100110111110001001101111100010011011000101111100010011011111001111101110011011111000100110111110001001101110110111010111001101111110011110001001101111100011101111101001111100010011011111000100110111000111011000100110111110001011010000110111110001001101111100010011011111100001101110011011111000100110111110001001101111100111000001001101110110111010111001101111100010011011111000100110111011100110000100110111110001011010000110111110111111011111010111110001001101111100010011011111001111101110011011111000100110111110001001101110110111010111001101111110011110001001101111100010011011111000100110111110001110110010110111110111111011111010111110000101100100110111110001001101111100010011011101110011000010011011111000100110111110001001101111110011110001001101111100010011011111000100110111110011100000100110111110001001101111100010011011111001111101110011011100011101100010011011111000101101000011011111000100110111110001001101111110000110111001101111100010011011111000100110111110011100000100110111011011101011100110111110001001101111100010011011101110011000010011011111000101101000011011111011111101111101011111000100110111110001001101111100111110111001101111100010011011111000100110111011011101011100110111111001111000100110111110001001101111100010011011111000111011001011011111011111101111101011111000010110010011011111000100110111110001001101110111001100001001101111100010011011111000100110111111001111000100110111110001001100000011111000001111101011111000100110111111111100011111000100110111110001001101111100010011011111000100110000001111100000111110001001101111100010011011000101111100010011011111000100110011000111111000100110111110001001101111100010011011100011101100010011011101110011000010011011111000100110111110001001101111110011110001001101111101001101111001101111100010011011111001110000010011011111000100110111110001011010000110111110111111011111010111110001001101111100010011011111001111101110011011111000100110111110001001101110110111010111001101111110011110001001101111100010011011111000100110111110001110110010110111110111111011111010111110000101100100110111110001001101111100010011011101110011000010011011111000100110111110001001101111110011110001001101111100010011000000111110000011111010111110001001101111111111000111110001001101111100010011011111000100110111110001001100000011111000001111100010011011111000100110110001011111000100110111110001001100110001111110001001101111100010011011111000100110111000111011000100110111011100110000100110111110001001101111100010011011111100111100010011011111010011011110011011111000100110111110011100000100110111110001001101111100010110100001101111101111110111110101111100010011011111000100110111110011111011100110111110001001101111100010011011101101110101110011011111100111100010011011111000100110111110001001101111100011101100101101111101111110111110101111100001011001001101111100010011011111000100110111011100110000100110111110001001101111100010011011111100111100010011011111010011011110011011111000100110111110001001101111100010001100110001111100001011001001101111100011101111100101111100010011011111000100110111110001110110010110111110001001101111100010011011111010011011110011011111000100110111110001001101110110111010111001101111100010011011111000100110111011100110000100110111110001011010000110111110111111011111010111110001001101111100010011011111001111101110011011111000100110111110001001101110110111010111001101111110011110001001101111100010011011111000100110111110001110110010110111110111111011111010111110000101100100110111110001001101111100010011011101110011000010011011111000100110111110001001101111110011110001001101111101001101111001101111100010011011111000100110111110001000110011000111110000101100100110111110001110111110010111110001001101111100010011011111000111011001011011111000100110111110001001101111101001101111001101111100010011011111010000111001100011100110001001101111100010000000110111001101111100010011011111000100110111110001001101111101000011100110001110011011111000100110111110000101100100110111110001001101111101110101110001101111100010011011111000100110111110001110111110010111110001110111110100111110001001101111100010011011111000100011111011011111000100110111000101111000100110111110001001111111000111110001001101111100010011011000001001100010011010001101001100010011011111000100110111110001001111100110111110001001101111100010011011111000111000011100011111000100011111011011111000100110111110001001101111100010011011111001110000010011010001101001100010011001100011111100010011011111000100110111110001110111110100111110001001101111100010011011111000100011111011011111000100110111110100001110011000111001100010011011111000100000001101110011011111000100110111110001001101111100010011011111010000111001100011100110111110001001101111100001011001001101111100010011011111011101011100011011111000100110111110001001101111100011101111100101111100011101111101001111100010011011111000100110111110001000111110110111110001001101111100010011011111000100110111110001001101111100010011011111000100110111110001001101111100010011011111000100110111110001001101111100010011011111000100110111110001001101111100010011011111000100110111110001001101111100010011011111000100110111110001001101111100010011011111000100110111110001001101111100010011011111000100110111110001001101111100010011011111000100110111110001001101111100010011011111000100110111110001001101111100010011011111000100110111110001001101111100010011011111000100110111110001001101111100010011011111000100110111110001001101111100010011011111000100110111110001001101111100010011011111000100110111110001001101111100010011011111000100110111110001001101111100010011011111000100110111110001001101111100010011011111000100110111110001001101111100010011011111000100110111110001001101111100010011011111000100110111110001001101111100010011011111000100110111110001001101111100010011011111000100110111110001001101111100010011011111000100110111110001001101111100010011011111000100110111110001001101111100010011011111000100110111110001001101111100010011011111000100110111110001001101111100010011011111000100110111110001001101111100010011011111000100110111110001001101111100010011011111000100110111110001001101111100010011011111000100110111110001001101111100010011011111000100110111110001001101111100010011011111000100110111110001001101111100010011011111000100110111110001001101111100010011011111000100110111110001001101111100010011011111000100110111110001001101111100010011011111000100110111110001001101111100010011011111000100110111110001001101111100010011011111000100110111110001001101111100010011011111000100110111110001001101111100010011011111000100110111110001001101111100010011011111000100110111110001001101111100010011011111000100110111110001001101111100010011011111000100110111110001001101111100010011011111000100110111110001001101111100010011011111000100110111110001001101111100010011011111000100110111110001001101111100010011011111000100110111110001001101111100010011011111000100110111110001001101111100010011011111000100110111110001001101111100010011011111000100110111110001001101111100010011011111000100110111110001001101111100010011011111000100110111110001001101111100010011011111000100110111110001001101111100010011011111000100110111110001001101111100010011011111000100110111110001001101111100010011011111000100110111110001001101111100010011011111000100110111110001001101111100010011011111000100110111110001001101111100010011011111000100110111110001001101111100010011011111000100110111110001001101111100010011011111000100110111110001001101111100010011011111000100110111110001001101111100010011011111000100110111110001001101111100010011011111000100110111110001001101111100010011011111000100110111110001001101111100010011011111000100110111110001001101111100010011011111000100110111110001001101111100010011011111000100110111110001001101111100010011011111000100110111110001001101111100010011011111000100110111110001001101111100010011011111000100110111110001001101111100010011011111000100110111110001001101111100010011011111000100110111110001001101111100010011011111000100110111110001001101111100010011011111000100110111110001001101111100010011011111000100110111110001001101111100010011011111000100110111110001001101111100010011011111000100110111110001001101111100010011011111000100110111110001001101111100010011011111000100110111110001001101111100010011011111000100110111110001001101111100010011011111000100110111110001001101111100010011011111000100110111110001001101111100010011011111000100110111110001001101111100010011011111000100110111110001001101111100010011011111000100110111110001001101111100010011011111000100110111110001001101111100010011011111000100110111110001001101111100010011011111000100110111110001001101111100010011011111000100110111110001001101111100010011011111000100110111110001001101111100010011011111000100110111110001001101111100010011011111000100110111110001001101111100010011011111000100110111110001001101111100010011011111000100110111110001001101111100010011011111000100110111110001001101111100010011011111000100110111110001001101111100010011011111000100110111110001001101111100010011011111000100110111110001001101111100010011011111000100110111110001001101111100010011011111000100110111110001001101111100010011011111000100110111110001001101111100010011011111000100110111110001001101111100010011011111000100110111110001001101111100010011011111000100110111110001001101111100010011011111000100110111110001001101111100010011011111000100110111110001001101111100010011011111000100110111110001001101111100010011011111000100110111110001001101111100010011011111000100110111110001001101111100010011011111000100110111110001001101111100010011011111000100110111110001001101111100010011011111000100110111110001001101111100010011011111000100110111110001001101111100010011011111000100110111110001001101111100010011011111000100110111110001001101111100010011011111000100110111110001001101111100010011011111000100110111110001001101111100010011011111000100110111110001001101111100010011011111000100110111110001001101111100010011011111000100110111110001001101111100010011011111000100110111110001001101111100010011011111000100110111110001001101111100010011011111000100110111110001001101111100010011011111000100110111110001001101111100010011011111000100110111110001001101111100010011011111000100110111110001001101111100010011011111000100110111110001001101111100010011011111000100110111110001001101111100010011011111000100110111110001001101111100010011011111000100110111110001001101111100010011011111000100110111110001001101111100010011011111000100110111110001001101111100010011011111000100110111110001001101111100010011011111000100110111110001001101111100010011011111000100110111110001001101111100010011011111000100110111110001001101111100010011011111000100110111110001001101111100010011011111000100110111110001001101111100010011011111000100110111110001001101111100010011011111000100110111110001001101111100010011011111000100110111110001001101111100010011011111000100110111110001001101111100010011011111000100110111110001001101111100010011011111000100110111110001001101111100010011011111000100110111110001001101111100010011011111000100110111110001001101111100010011011111000100110111110001001101111100010011011111000100110111110001001101111100010011011111000100110111110001001101111100010011011111000100110111110001001101111100010011011111000100110111110001001101111100010011011111000100110111110001001101111100010011011111000100110111110001001101111100010011011111000100110111110001001101111100010011011111000100110111110001001101111100010011011111000100110111110001001101111100010011011111000100110111110001001101111100010011011111000100110111110001001101111100010011011111000100110111110001001101111100010011011111000100110111110001001101111100010011011111000100110111110001001101111100010011011111000100110111110001001101111100010011011111000100110111110001001101111100010011011111000100110111110001001101111100010011011111000100110111110001001101111100010011011111000100110111110001001101111100010011011111000100110111110001001101111100010011011111000100110111110001001101111100010011011111000100110111110001001101111100010011011111000100110111110001001101111100010011011111000100110111110001001101111100010011011111000100110}
$
}

\begin{table}[th]
\centering
\begin{tabular}{ |l|l| }
  \hline
  \multicolumn{2}{|c|}{machines} \\
  \hline
C hello world & $2.19 \times10^{-191}$\% \\
1D Fredkin gate ECAM rule 22 & $2.36 \times 10^{-620}$\% \\
Cyclic tag system ECA rule 110 & $1.18 \times 10^{-16928}$\% \\
  \hline
\end{tabular}
\caption{Probability to get an algorithm systematically or randomly.}
\label{ra_tbl1}
\end{table}

\section{Final note}

In the aftermath of an apocalyptic scenario, the prospect of recovering and constructing algorithms remains, although the likelihood of stumbling upon them either randomly or through systematic means is consistently low. Throughout history, we often observe that transformative events akin to an apocalypse are necessary for significant advancements to take place.

\bibliographystyle{ws-rv-van}
\bibliography{ws-rv-sample}

\end{document}